\newif\ifemulate
\emulatetrue  % comment this out for AAS preprint
\pdfoutput=1

\ifemulate
\documentclass{emulateapj}
\usepackage{apjfonts}
\else
\documentclass[preprint]{aastex}
\fi

\usepackage[pagebackref=false]{hyperref}
\hypersetup{
  bookmarks=true,         % show bookmarks bar
  bookmarksopen=true,     % expand bookmark tree on open
  bookmarksnumbered=true, % have section number in bookmarks
  hyperfootnotes=true,    % footnote marks are links to footnote text
  colorlinks=true,        % false: boxed links; true: colored links
  linkcolor=blue,         % color of internal links
  citecolor=blue,         % color of links to bibliography
  urlcolor=blue,          % color of external links
  pdffitwindow=false,     % window fit to page when opened
  pdfstartview={FitH},    % fits the width of the page to the window
  pdftitle={On the 10-micron silicate feature in Active Galactic Nuclei},
  pdfauthor={Nikutta et al.}
}

\slugcomment{Version of \today}

\newcommand{\E}[1]{\hbox{$10^{#1}$}}

\newcommand{\tk}[1]{\tablenotemark{#1}}
\newcommand\about {\hbox{$\sim$}}
\newcommand\C     {\hbox{\textsc{Clumpy}}}
\renewcommand\deg {\hbox{$^\circ$}}
\newcommand\Emin  {\hbox{$E_{\rm min}$}}
\newcommand\Fcont {\hbox{$F_{\rm cont}$}}
\newcommand\Lo    {\hbox{$L_{\odot}$}}
\newcommand\Ltw   {\hbox{$L_{12}$}}
\newcommand\mic   {\hbox{$\mu$m}}
\newcommand\Pesc  {\hbox{$P_{\rm esc}$}}
\newcommand\Pobs  {\hbox{$P_{\rm obsc}$}}
\newcommand\pg    {\hbox{PG1211+143}}
\newcommand\sst   {\hbox{SST1721+6012}}
\newcommand\sten  {\hbox{$S_{10}$}}
\newcommand\tv    {\hbox{$\tau_V$}}
\newcommand\q     {\hbox{$q$}}
\newcommand\No    {\hbox{$N_0$}}
\newcommand\sig   {\hbox{$\sigma$}}
\renewcommand\i   {\hbox{$i$}}
\newcommand\Y     {\hbox{$Y$}}

\defcitealias{Nenkova+2002}{N02}
\defcitealias{Nenkova+2008a}{N08a}
\defcitealias{Nenkova+2008b}{N08b}
\defcitealias{OHM1992}{OHM}

\shorttitle{10-micron feature in AGN}
\shortauthors{Nikutta et al.}

\begin{document}

\title{On the 10-micron silicate feature in Active Galactic Nuclei}

\author{Robert Nikutta\altaffilmark{1},
        Moshe Elitzur\altaffilmark{1}
        and Mark Lacy\altaffilmark{2}}

\altaffiltext{1}{Department of Physics and Astronomy, University
                 of Kentucky, Lexington, KY 40506-0055;
                 robert@pa.uky.edu, moshe@pa.uky.edu}
\altaffiltext{2}{Spitzer Science Center, California Institute of
                 Technology, Pasadena, CA 91125, USA;
                 mlacy@ipac.caltech.edu}

\begin{abstract}

  The 10\mic\ silicate feature observed with Spitzer in active
  galactic nuclei (AGN) reveals some puzzling behavior. It (1) has
  been detected in emission in type~2 sources, (2) shows broad,
  flat-topped emission peaks shifted toward long wavelengths in
  several type~1 sources, and (3) is not seen in deep absorption in
  any source observed so far. We solve all three puzzles with our
  clumpy dust radiative transfer formalism. Addressing (1), we present
  the spectral energy distribution (SED) of \sst, the first type~2
  quasar observed to show a clear 10\mic\ silicate feature in
  emission. Such emission arises in models of the AGN torus easily
  when its clumpy nature is taken into account. We constructed a large
  database of clumpy torus models and performed extensive fitting of
  the observed SED. We find that the cloud radial distribution varies
  as $r^{-1.5}$ and the torus contains 2--4 clouds along radial
  equatorial rays, each with optical depth at visual \about
  60--80. The source bolometric luminosity is \about
  $3\cdot\E{12}\,\Lo$. Our modeling suggests that $\la\,35\%$ of
  objects with tori sharing these characteristics and geometry would
  have their central engines obscured. This relatively low obscuration
  probability can explain the clear appearance of the 10\mic\ emission
  feature in \sst\ together with its rarity among other
  QSO2. Investigating (2) we also fitted the SED of \pg, one of the
  first type~1 QSOs with a 10\mic\ silicate feature detected in
  emission.  Together with other similar sources, this QSO appears to
  display an unusually broadened feature whose peak is shifted toward
  longer wavelengths. Although this led to suggestions of non-standard
  dust chemistry in these sources, our analysis fits such SEDs with
  standard galactic dust; the apparent peak shifts arise from simple
  radiative transfer effects. Regarding (3) we find additionally that
  the distribution of silicate feature strengths among clumpy torus
  models closely resembles the observed distribution, and the feature
  never occurs deeply absorbed. Comparing such distributions in
  several AGN samples we also show that the silicate emission feature
  becomes stronger in the transition from Seyfert to quasar
  luminosities.

\end{abstract}


\section{INTRODUCTION}

Unified schemes of active galactic nuclei (AGN) require an obscuring
dusty torus around the central source, giving rise to a type~1 line
spectrum when there is direct view of the central engine and type~2
characteristics when it is blocked \citep[e.g.][]{Antonucci1993,
  UrryPadovani1995}. The torus, which is comprised of dusty clouds
that are individually optically thick \citep{KrolikBegelman1988},
reprocesses the radiation it absorbs into longer wavelengths, creating
a distinct signature in the observed infrared. Silicates, a major
constituent of astronomical dust, reveal their presence through the
spectral feature at 10~\mic. Among type~1 AGN, QSOs display the
feature in emission \citep{Siebenmorgen+2005,Hao+2005,Sturm+2005},
while average SEDs of Seyfert~1 galaxies have either a flat 10\mic\
feature \citep{Wu+2009} or show it in mild absorption
\citep{Hao+2007}. Seyfert~2 galaxies generally display an absorption
feature with limited depth, much shallower than in ultra-luminous IR
galaxies \citep[e.g.][]{Hao+2007,Levenson+2007}. An intriguing result
comes from the Spitzer observations of seven high-luminosity type~2
QSOs by \citet{Sturm+2006}. While individual spectra appear
featureless, the sample average spectrum shows the 10\mic\ feature in
{\em emission}.

Heated dust will produce the feature in emission whenever it is
optically thin. When the dust optical depth at 10~\mic\ exceeds unity,
the feature still appears in emission in viewing of the illuminated
face of the dust but in absorption when the dust is between the
observer and heating source. In the absence of a formalism for
radiative transfer in clumpy media, early models of the AGN torus
employed smooth density distributions instead
\citep[e.g.][]{PierKrolik1992, PierKrolik1993, GranatoDanese1994,
  EfstathiouRowan-Robinson1995, Granato+1997, Fritz+2006}. These
models predict that type~1 sources, where the observer has a direct
view of the torus inner, heated face, will generally produce an
emission feature, although some examples of absorption features do
exist \citep[]{PierKrolik1992, EfstathiouRowan-Robinson1995}. Type~2
viewing in most cases produces an absorption feature, whose depth is
quite large on occasion, much larger than ever observed. An emission
feature is rarely produced from such viewing, but it should be noted
that it is under certain conditions for instance in the model
presented by \citet{Fritz+2006}. A formalism for handling clumpy media
was developed by \citet{Nenkova+2002,Nenkova+2008a} (hereafter N02 \&
N08a); the formalism holds for volume filling factors as large as
10\%. Their models show that a clumpy torus will never produce a very
deep absorption feature and that the feature displays a much richer
behavior than in smooth density models; in particular, type~1 viewing
can produce an absorption feature in certain models and type~2 viewing
can lead to an emission feature in others \citep[N08b
henceforth]{Nenkova+2008b}.

While the \cite{Sturm+2006} data suggest the possibility of a 10\mic\
emission feature in QSO2, the only unambiguous evidence for such a
feature in a type~2 AGN was presented recently for the Seyfert galaxy
NGC~2110 \citep{Mason+2009}.\footnote{\cite{Teplitz+2006} have
  suggested a 10\mic\ emission feature in the Spitzer spectrum of QSO2
  FSC10214+4724. The suggestion is problematic because the object's
  redshift is so high ($z$ = 2.2856) that the 10\mic\ feature was not
  fully in the spectral range of the IRS instrument. The rest-frame
  spectrum is cut off around 12\mic, before the continuum longward of
  the feature could be established.} Here we present the first
unambiguous case of an emission feature in a type~2 quasar, \sst, and
perform extensive fitting of its spectral energy distribution (SED)
with clumpy torus models.

The comparison of torus model predictions with observations is
somewhat problematic because the overwhelming majority of these
observations do not properly isolate the torus IR emission. Starburst
emission is a well known contaminant in many cases, and we selected
\sst\ for modeling precisely for this reason as its spectrum seems
free of starburst indicators. However, even IR from the immediate
vicinity of the AGN may not always originate exclusively from the
torus. High-resolution observations of NGC1068 by \cite{Cameron+1993}
and recently by \cite{Mason+2006} demonstrate that the torus
contributes less than 30\% of the 10\mic\ flux collected with
apertures $\ge 1''$ in this object, with the bulk of this flux coming
from dust in the ionization cones (\cite{Braatz+1993} also found that
at least 40\% of the 12.4\mic\ flux in this source do not originate
from the torus). The significance of IR emission from the narrow line
region (NLR) was noted also by \citet{Schweitzer+2008}. However,
because the dust in the ionization cones is optically thin, its IR
emission is isotropic and does not generate differences between types
1 and 2. Observations show that such differences do exist. In
particular, the \cite{Hao+2007} compilation of \emph{Spitzer} IR
observations shows a markedly different behavior for the 10\mic\
feature between Seyfert 1 and 2 galaxies. Accepting the framework of
the unification scheme, these differences can be attributed only to
the torus contribution. Thus it seems that, unfortunately, a general
rule does not exist and the situation must be investigated case by
case. Our aim here is to examine whether the torus contribution alone
can reproduce the observed SED of \sst, yielding a range of possible
parameter values that describe the dusty cloud distribution in this
source (\S\ref{sec:sil_emission_qso2}).

We also investigate the cause for apparent shifts of the silicate
feature peaks towards long wavelengths (\S\ref{sec:peakshift}). Such
shifts have been reported for sources that show the 10\mic\ feature in
emission \citep{Siebenmorgen+2005,Sturm+2005,Hao+2005}, and attributed
to non-standard dust chemistry. However, these shifts were never seen
in absorption, suggestive of radiative transfer effects
instead. Finally, in \S\ref{sec:s10} we compare the observed
distribution of silicate feature strengths among the \cite{Hao+2007}
sample of AGN with the synthetic distribution of feature strengths in
our database of clumpy torus model SEDs.

\section{SILICATE 10-MICRON EMISSION FEATURE IN QSO2}
\label{sec:sil_emission_qso2}

Although not expected in type~2 sources, possible detection of the
10\mic\ emission feature was reported by \citet{Sturm+2006}. The
feature was only identified after averaging the SEDs of a number of
type~2 QSOs, which individually show no significant indication of the
feature. Recently \citet{Mason+2009} presented the first unequivocal
detection of an emission feature in an individual type~2 source, the
Seyfert galaxy NGC~2110. We present the {\em Spitzer} SED of the
type~2 quasar \sst\ that shows the 10\mic\ and 18\mic\ silicate
features in emission. In this section we report on the results of
fitting the SED of \sst\ with clumpy torus models, and derive multiple
parameters characterizing the source.

\subsection{Observations}
\label{sec:observations}

The source SSTXFLS J172123.1+601214 was first identified as an AGN
candidate in the Spitzer First Look Survey (FLS) by
\citet{Lacy+2004}. It has a redshift of $z = 0.325$, and was not
present in the SDSS at that time. In 2007, \citet{Lacy+2007a}
categorized it as a type~2 quasar based on the presence of optical,
narrow [\ion{N}{5}] emission lines and through emission line ratio
diagnostics introduced by \citet{Baldwin+1981}. In the same year
\citet{Lacy+2007b} presented, together with other sources, a
wide-range SED for this source, including a mid-IR spectrum taken by
the Infrared Spectrograph (IRS) aboard Spitzer.

The IRS observations (Astronomical Observation Request 1406768) were
taken on 2005 August 14 in staring mode using the short and long low
resolution modules to obtain continuous coverage from 5.2--38~\mic,
and were passed through the S14.0 version of the SSC pipeline. The
signal-to-noise ratio varied through the spectrum, the deepest
observations being targeted on the redshifted wavelengths of the
strong spectral features expected to lie in the 7--15~\mic\ range. In
the short wavelength module, two 14s ramps were taken in second order,
and a single 60s ramp in first order. In the long wavelength module,
two 30s ramps were taken in both first and second order. The spectra
from each module were optimally extracted using
SPICE.\footnote{\url{http://ssc.spitzer.caltech.edu/postbcd/spice.html}}
The resulting spectra were trimmed, combined, and resampled in
constant energy bins of $\Delta\lambda / \lambda \approx 0.01$,
resulting in a spectrum ranging from 4.0~\mic\ to 27.1~\mic\ (rest
wavelength). Uncertainty estimates from SPICE were propagated through
the process in the usual manner. For the fitting we excluded a few
data points at shorter and at longer wavelengths due to poor
signal-to-noise ratio. Additionally, we make use of two photometric
data points from the Infrared Array Camera (IRAC) component of the
{\em Spitzer} First Look Survey \citep{Lacy+2005} at rest wavelengths
of 2.7~\mic\ and 3.4~\mic, both with very small intrinsic
uncertainties, as they greatly help defining the shape of the SED in
the regions of hot dust emission. Cross-calibration between IRS and
IRAC is accurate to better than 10\% (L. Yan, personal communication).

Despite a certain noisiness in the IRS spectrum, a clear presence of
polycyclic aromatic hydrocarbon features (PAHs) can be safely
excluded. Considering additionally its lack of a [\ion{Ne}{2}]
emission line at 12.8~\mic, the spectrum shows no signs of star
formation. Furthermore, the SED seems free of other emission lines,
with one disputable exception. Locally, the flux peaks around
10.5~\mic, which coincides with the [\ion{S}{4}] emission line at
10.51~\mic\ reported to be found in 11 out of 12 type~2 sources by
\citet{Zakamska+2008}. This radiation, if indeed credited with an
emission line, would stem from the AGN itself, but our spectrum does
not show any other lines originating from the AGN, like [\ion{Ne}{3}]
at 15.5~\mic\ and [\ion{Ne}{5}] at 14.3~\mic. Within the frame of this
work, we therefore attribute the peak flux at \about 10.5~\mic\
entirely to silicate emission.

\subsection{Modeling}
\label{sec:modeling}

\citetalias{Nenkova+2002} \& \citetalias{Nenkova+2008a} describe an
analytic formulation of radiative transfer in a clumpy, dusty medium
heated by a radiation source. The formalism was implemented in the
code \C, which takes as input a toroidal distribution of point-like
dust clouds around a central source. The dust in each individual cloud
has an optical depth $\tv$, defined at 0.55~\mic, and standard ISM
composition of 47\% graphite with optical constants from
\citet{Draine2003} and 53\% ``cold'' silicates from \citet*{OHM1992}
(OHM hereafter). The dust sublimation temperature defines the torus
inner radius $R_d$ and is set to $1500\, {\rm K}$. The cloud
distribution is parametrized with the radial power law $1/r^q$ between
$R_d$ and the outer radius $Y R_d$, where \q\ and \Y\ are free
parameters. Another free parameter is \No, the average number of
clouds along a radial equatorial ray. In polar direction the number of
clouds per radial ray is characterized by a Gaussian, so that at angle
$\beta$ from the equatorial plane it is $N_0\,e^{-(\beta/\sigma)^2}$,
with \sig\ the last free parameter of the cloud distribution. The
final parameter is \i, the observer's viewing angle measured from the
torus axis.

We employed \C\ to produce a large database\footnote{Models are
  available at \url{http://www.pa.uky.edu/clumpy}} of model SEDs
$f_\lambda = \lambda F_\lambda / F_{AGN}$, with $F_{AGN}$ the total
bolometric flux. The observations provide a set of fluxes, $F^o_j$, at
wavelengths $\lambda_j$, $j = 1\ldots N$. Our fitting procedure
involves searching the entire database for the model that minimizes
the error
\begin{equation}
  \label{eq:E}
  E = \frac{1}{N} \sqrt{ \sum_{j=1}^N
      \left( \frac{F_{AGN} \cdot f^m_j - \lambda_j F^o_j}{\Delta_j} \right)^2}\,,
\end{equation}
where $\Delta_j$ are individual errors on the $\lambda_j F^o_j$, and
$f^m_j$ are the model fluxes at the same set of wavelengths as the
data. Each model SED is scaled by the factor $F_{AGN}$ that minimizes
$E$, determining the AGN bolometric flux for this model. Since the
data dynamic range is only $\approx 3$, the fitting procedure can be
safely executed in linear space.

\subsection{Results}
\label{sec:fittingresults}

%%%%%%%%%%%%%%%%%%%%%%%%%%%%%%%%%%%%%%%%%%%%%%%%%%%%%%%%%%%%%%%
\begin{deluxetable}{cl}
\tablewidth{\columnwidth}
\tablecaption{\textsc{Clumpy} parameters used in fitting \label{tab:clumpyparams}}
\tablehead{\colhead{Parameter} &\multicolumn{1}{l}{Sampled Values}}
\startdata
\q   & 0, 0.5, 1, 1.5, 2, 2.5, 3 \\
\No  & 1 - 25 \\
\tv  & 5, 10, 20, 30, 40, 60, 80, 100, 150, 200, 300, 500 \\
\sig & 15, 20, 25, 30, 35, 40, 45, 50, 55, 60, 65, 70, 75, 80 \\
\i   & 0, 10, 20, 30, 40, 50, 60, 70, 80, 90 \\
\Y   & 2 - 5, 10, 20, 30, 40, 50, 60, 70, 80, 90, 100, 150, 200
\enddata
\tablecomments{\sig\ and \i\ are measured in degrees.}
\end{deluxetable}
%%%%%%%%%%%%%%%%%%%%%%%%%%%%%%%%%%%%%%%%%%%%%%%%%%%%%%%%%%%%%%%

\begin{figure}
\includegraphics[width=\columnwidth]{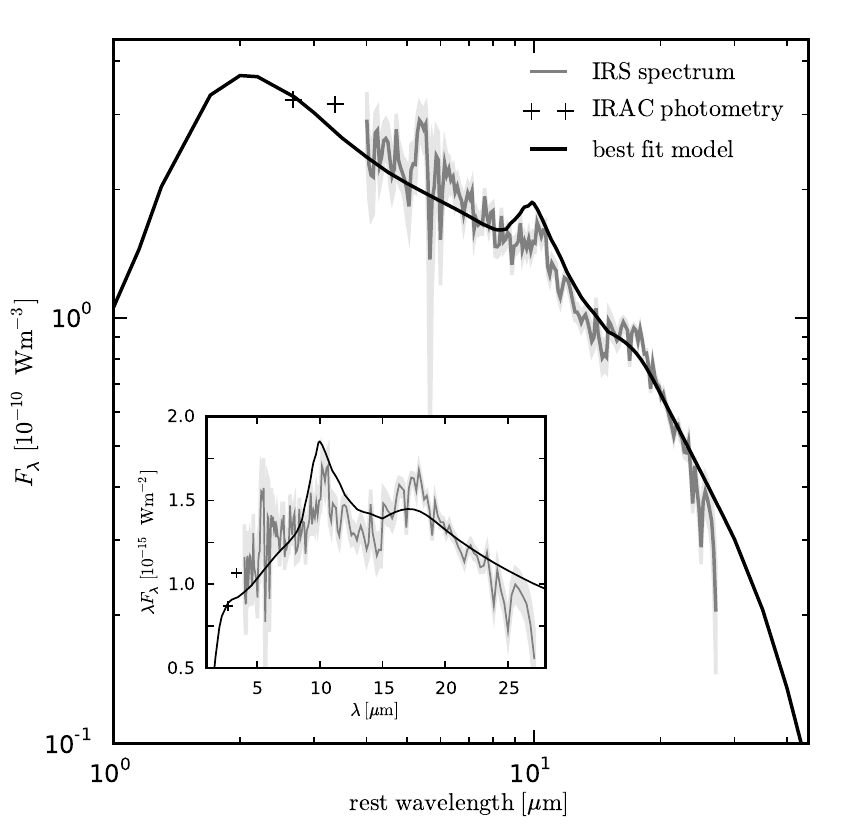}
\caption{SED of \sst. Spitzer IRS data are shown in dark gray with the
  errors in light gray shade. Two IRAC photometry points are marked
  with crosses. The black line shows the best-fit \C\ model, which
  produces an error \Emin\ = 0.212 (see~Equation~\ref{eq:E}). Its
  parameters are $q=1.5$, $N_0=3$, $\tv=80$, $\sigma=20$, $Y=30$ and
  $i=60$. The inset shows the data and the best fit model using
  $\lambda F_\lambda$ and linear scales for a better display of the
  10\mic\ emission feature.}
\label{fig:f1}
\end{figure}

We calculated $E$ for all the \C\ models whose parameters are listed
in Table~\ref{tab:clumpyparams}, resulting in a database of more than
4.7 million entries. This large set contains as a subset all the
parameters that \citetalias{Nenkova+2008b} found to be
plausible. Figure~\ref{fig:f1} shows the data and the best-fitting \C\
model. The two photometric IRAC points play a crucial role in the fits
by expanding the data into the short wavelengths.

%%%%%%%%%%%%%%%%%%%%%%%%%%%%%%%%%%%%%%%%%%%%%%%%%%%%%%%%%%%%%%%
%%%%%%%%%%%%%%%%%%%%%%%%%%%%%%%%%%%%%%%%%%%%%%%%%%%%%%%%%%%%%%%

\begin{figure}
\includegraphics[width=\columnwidth]{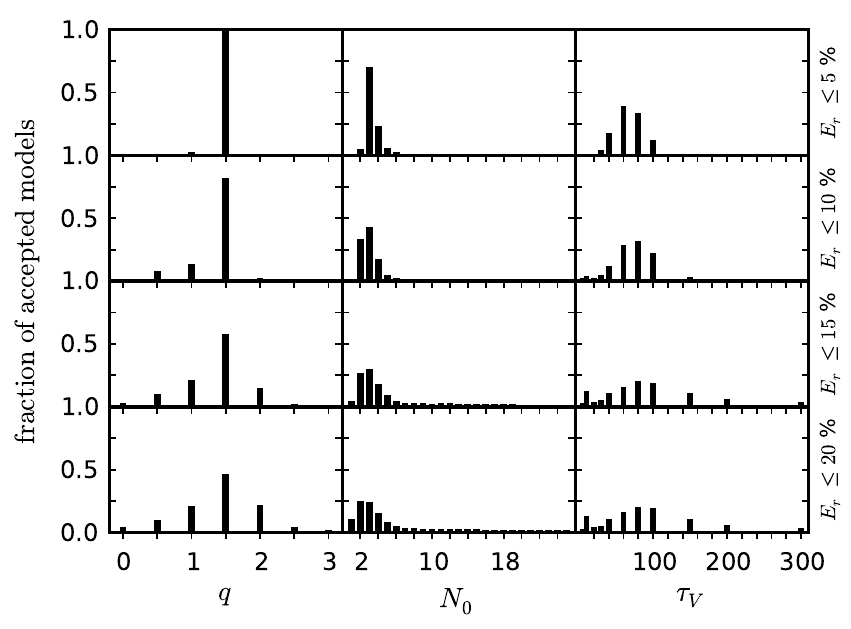}
\caption{Bar diagrams of three \C\ parameters well constrained by
  fitting. From left to right, the columns correspond to \q, \No, and
  \tv. The parameters were sampled as listed in
  \mbox{Table~\ref{tab:clumpyparams}}. Rows correspond, from top to
  bottom, to an increasing acceptance on the fitting error relative to
  the best-fit model, as marked on the right, with the resulting
  number of models increasing accordingly --- 199, 1691, 5210 and
  12854. The height of the bar at any value of a parameter is the
  fraction of all accepted models.}
\label{fig:f2}
\end{figure}

\begin{figure}
\includegraphics[width=\columnwidth]{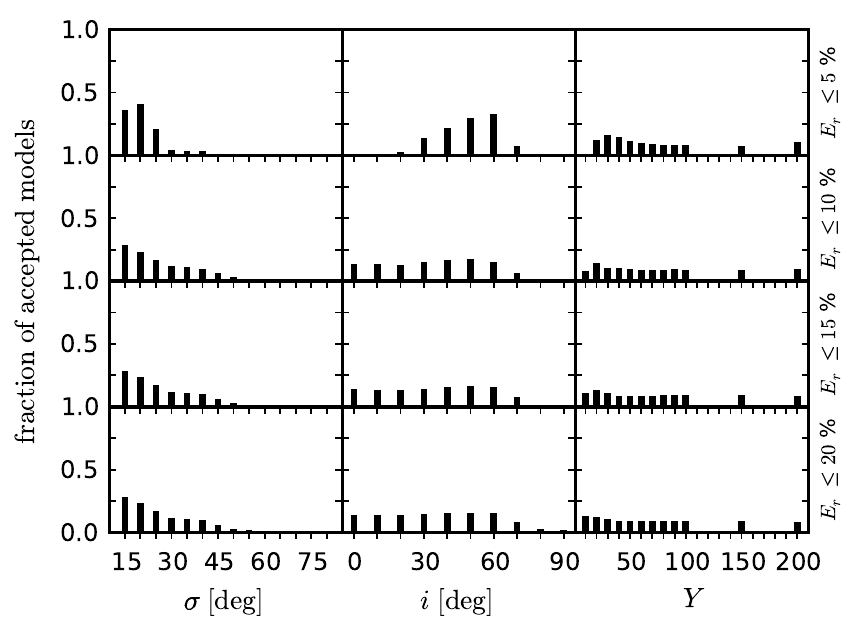}
\caption{Same as Figure~\ref{fig:f2}, but for the three less
  well constrained \C\ parameters \sig, \i, and \Y. These
  distributions flatten out more quickly with growing acceptance
  error.}
\label{fig:f3}
\end{figure}

Although the model presented in Figure~\ref{fig:f1} produces the
smallest nominal error $E$, a number of other models have errors that
differ from it only in the third significant digit. Because of the
large degeneracy of the radiative transfer problem for heated dust,
the SED is a poor constraint on the properties of the source; a
meaningful determination of model parameters requires also
high-resolution imaging at various wavelengths
\citep[e.g.,][]{Vinkovic+2003}. The axially symmetric clumpy torus
model requires a relatively large number of input parameters, further
exacerbating the degeneracy problem. We define $E_r =
100\cdot(E-\Emin)/\Emin$ as the relative deviation of a model from the
best-fit one. Then, 199 models have $E_r \le 5\%$, within a fraction
of the minimal error \Emin\ = 0.212, and the bar diagrams of these
models are shown in the top rows of Figures~\ref{fig:f2} and
\ref{fig:f3} for each of the six parameters. All but two of these
models share the same value of $q = 1.5$, indicating that this
parameter can be considered well constrained. Similarly, for 90\% of
all models \No\ is either 3 or 4, so this parameter is only slightly
less well constrained. The distributions of the parameters \tv, \sig\
and \i\ are broader, but still show well defined peaks. For these
parameters we can only deduce a plausible range. In contrast, the
parameter \Y\ has a flat distribution that covers every sampled value
$Y \ge 20$; this parameter is undetermined, except for the indication
of a lower bound.

%%%%%%%%%%%%%%%%%%%%%%%%%%%%%%%%%%%%%%%%%%%%%%%%%%%%%%%%%%%%%%%
%%%%%%%%%%%%%%%%%%%%%%%%%%%%%%%%%%%%%%%%%%%%%%%%%%%%%%%%%%%%%%%

%%%%%%%%%%%%%%%%%%%%%%%%%%%%%%%%%%%%%%%%%%%%%%%%%%%%%%%%%%%%%%%
%%%%%%%%%%%%%%%%%%%%%%%%%%%%%%%%%%%%%%%%%%%%%%%%%%%%%%%%%%%%%%%

The choice $E_r \le 5\%$ is of course arbitrary. Increasing slightly
the range of accepted models, the bar diagrams can be expected to
remain peaked if the parameters are well constrained. The figures show
that this is indeed the case for \q\ and \No, whose distributions
remain reasonably peaked in spite of the large increase in the number
of accepted models (almost 13,000 are selected by the criterion $E_r
\le 20\%$). To a lesser degree, this is also the case for \tv. In
contrast, the distributions of \sig\ and \i, which also start out
peaked, flatten out significantly as the acceptance criterion is
relaxed, indicating that SED analysis lacks the predictive power to
constrain these parameters in \sst. The only meaningful results are
that, in all likelihood, $\sigma \la 50\deg$ and $i \le 70\deg$, i.e.,
edge-on viewing is excluded. Furthermore, these parameters are not
entirely independent of each other since a clear line of sight to the
AGN can be obtained for different combinations of the two. In fact,
the interdependence of \sig, \i, and to some degree \No\, constitutes
the greatest source of degeneracy within the clumpy torus SEDs. The
final parameter, the torus radial thickness \Y, is undetermined. As
noted already in \citetalias{Nenkova+2008b}, the SEDs of models with a
steep radial cloud distribution ($q > 1$) are insensitive to
increasing \Y\ because most of the clouds are concentrated in the
torus inner region. The only constraint we can deduce is the lower
bound $Y \ge 10$, indicating that the torus could be compact, in
agreement with other AGN observations (see \citetalias{Nenkova+2008b}
and references therein).

Table~\ref{tab:parametervaluesSST} summarizes the likely values
constrained by fitting. We cannot give exact confidence intervals
since our distributions are not continuous. If a parameter is
perfectly constrained, all models then have the same value. Denoting
by $H$ the fraction of models at the distribution peak value, such a
parameter would have $H = 1$. On the other hand, a flat distribution
over the entire range of sampled values indicates a completely
non-constrained parameter. If the number of parameter values in the
sampled range is $B$, the height of each bar would then be
$1/B$. Introduce $w = H/B$. A perfectly constrained parameter will
have $w = 1\ (= H = B)$ while for an unconstrained parameter $w =
1/B^2$, decreasing when the number of sampled values is increasing. We
select as our sample the 1691 models with $E_r \le 10\%$. While
admittedly arbitrary, this selection ensures a strict acceptance
criterion while still giving a statistically large sample. For each of
the model parameters we identify the minimal interval around the
distribution peak containing at least 90\% of the sample's
models. These ranges are listed in Table~\ref{tab:parametervaluesSST},
together with the number of sampled values (bars) in these intervals,
which is our measure of $B$. The last column lists the corresponding
values of $w$, reinforcing the perception conveyed by the bar diagrams
regarding the degree of confidence (or lack thereof) in each of the
derived model parameters.

This analysis shows that the radial cloud distribution in \sst\ is
well constrained at $q = 1.5$; although the 90\% range contains also
$q = 1$, 81\% of the models are at $q = 1.5$. The likely value of \No\
is similarly well constrained to the range 2--4; even though this
parameter was densely sampled in steps of 1 all the way to 25, half of
the 50,000 best models fall within this narrow range. The third
reasonably well determined parameter is $\tv \approx 80$, whose likely
value is between 30 and 100. Note that the values of these parameters
for the best-fitting model are $q = 1.5$, $N_0 = 3$ and \mbox{$\tv =
  80$}, and that the close agreement with the distribution peaks is
not a given --- in principle, the best-fit model could fall anywhere
inside the acceptable ranges. We have tried to put stronger
constraints on the less well-defined parameters \sig, \i, and \Y, by
holding the values of the relatively well-constrained parameters fixed
at $q = 1.5$, $N_0 = 2-4$, and \mbox{$\tv = 60-100$}. This had little
effect on the distributions of the unconstrained parameters, although
the \sig\ bar-diagrams became slightly more peaked, showing a hint of
greater preference for $\sigma \approx 15-30$. We conclude that it is
impossible to deduce \sig, \i, and \Y\ for \sst\ from SED analysis
alone.

%%%%%%%%%%%%%%%%%%%%%%%%%%%%%%%%%%%%%%%%%%%%%%%%%%%%%%%%%%%%%%%%%%
\begin{deluxetable}{ccccccc}
\tablewidth{\columnwidth}
\tablecaption{Properties of fitted parameters for \sst \label{tab:parametervaluesSST}}
\tablehead{\colhead{Parameter} &\colhead{Best Fit} &\colhead{Peak\tk{a}} &\colhead{90\%-Range\tk{b}} &\colhead{B\tk{c}} &\colhead{H\tk{d}} &\colhead{w\tk{e}}}
\startdata
\q     & 1.5 & 1.5 & \phd 1 -- 1.5       & 2  & 0.81 & 0.40 \\
\No    & 3   & 3   & \phn 2 -- 4\phn\phn & 3  & 0.42 & 0.14 \\
\tv    & 80  & 80  &     30 -- 100       & 5  & 0.30 & 0.06 \\
\Y     & 30  & 20  &     20 -- 200       & 11 & 0.13 & 0.01 \\
\hline
\sig   & 20  & 15  &     15 -- 40\phn    & 6  & 0.27 & 0.05 \\
\i     & 60  & 50  & \phn 0 -- 60\phn    & 7  & 0.16 & 0.02
\enddata
\tablecomments{Statistical indicators for the sample of all 1691 models with $E_r \leq 10\%$ deviation from the best-fit model.}
\tablenotetext{a}{Value of the parameter at the distribution peak.}
\tablenotetext{b}{Range around the peak containing at least 90\% of the sample models.}
\tablenotetext{c}{Number of sampled values in the 90\%-range.}
\tablenotetext{d}{Fraction of all accepted models at the distribution peak.}
\tablenotetext{e}{Measure of how well the parameter is constrained (see text);
  the closer $w$ is to unity the higher is the significance of the determined
  value. The values for the last two entries cannot be directly compared to all
  others since the range of both \sig\ and \i\ is finite whereas for all other
  parameters it is in principle unlimited.}
\end{deluxetable}
%%%%%%%%%%%%%%%%%%%%%%%%%%%%%%%%%%%%%%%%%%%%%%%%%%%%%%%%%%%%%%%%%%%%

\subsection{Source Type}
\label{sec:source-type}

In the standard form of the unification approach, the classification
of an AGN as type 1 or 2 is uniquely determined by the relation
between the viewing angle \i\ and the torus angular thickness \sig. In
a clumpy medium, on the other hand, the source type is a matter of
probability. Denote by $N(i)$ the average number of clouds along a
radial ray at angle \i, then
\begin{equation}
  \label{eq:Pesc}
  \Pesc(i) = e^{-N(i)}
\end{equation}
is the probability that a photon emitted by the AGN will escape the
torus. The source has a probability $\Pesc(i)$ to appear as a type 1
AGN and $\Pobs(i) = 1 - \Pesc(i)$ as a type 2. With our Gaussian
parametrization for the cloud angular distribution,
\begin{equation}
  \label{eq:Ni}
  N(i) = N_0\,e^{\,-\,[(90-i)/\sigma]^2} \,.
\end{equation}
The AGN type is probabilistic, and it depends on \i, \sig\ and \No.

\begin{figure}
\includegraphics[width=\columnwidth]{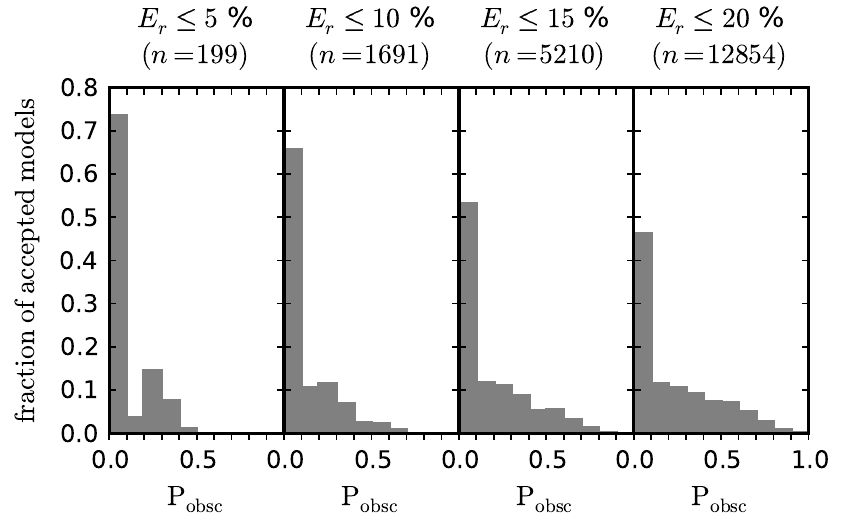}
\caption{Histograms of $\Pobs$, the probability that the AGN is
  obscured by the torus in the \C\ model (the probability that the
  model produces a type~2 source). Each panel corresponds to a
  different maximal acceptance error, as marked at the top together
  with the corresponding number of models. Each bin width is 0.1 and
  its height is the fraction of all accepted models. The mean value in
  each panel, from left to right, is 0.10, 0.11, 0.17 and 0.22}.
\label{fig:f4}
\end{figure}

Since \sst\ is a type~2 quasar, the a priori expectation would be that
\Pobs\ is large. We find this not to be the case. The best-fit model
has \mbox{\Pobs\ = 27\%}, and more than 75\% of all models with $E_r
\le 15\%$ have \mbox{$\Pobs \le 33\%$}. Figure~\ref{fig:f4} displays
the histograms of \Pobs\ for the models accepted at various tolerance
levels, showing that the majority of models have \mbox{$\Pobs \le
  10\%$} (in the first 3 panels). Such low probability would pose a
problem if these were the numbers for a large sample of type~2
sources. However, \sst\ is a relatively rare type~2 quasar with a
clear 10\mic\ emission feature; of the more than twenty QSO2 with
measured IR SEDs, NGC~2110 is the only other source with such
unambiguous emission feature. The emission feature requires a direct
line of sight to a significant fraction of the hot surfaces of
directly illuminated clouds on the far inner side of the
torus. Because obscuration of the AGN involves a single line of sight
while the IR flux measurements integrate over many lines of sight, the
relatively low values of \Pobs\ that emerge from the modeling are
commensurate with the clear appearance of the 10\mic\ emission feature
in \sst\ and its rarity among other QSO2 (see also
\S\ref{sec:peakshift}).

%%%%%%%%%%%%%%%%%%%%%%%%%%%%%%%%%%%%%%%%%%%%%%%%%%%%%%%%%%%%%%%
%%%%%%%%%%%%%%%%%%%%%%%%%%%%%%%%%%%%%%%%%%%%%%%%%%%%%%%%%%%%%%%

\subsection{AGN Luminosity}
\label{sec:LAGN}

\begin{figure}
\includegraphics[width=\columnwidth]{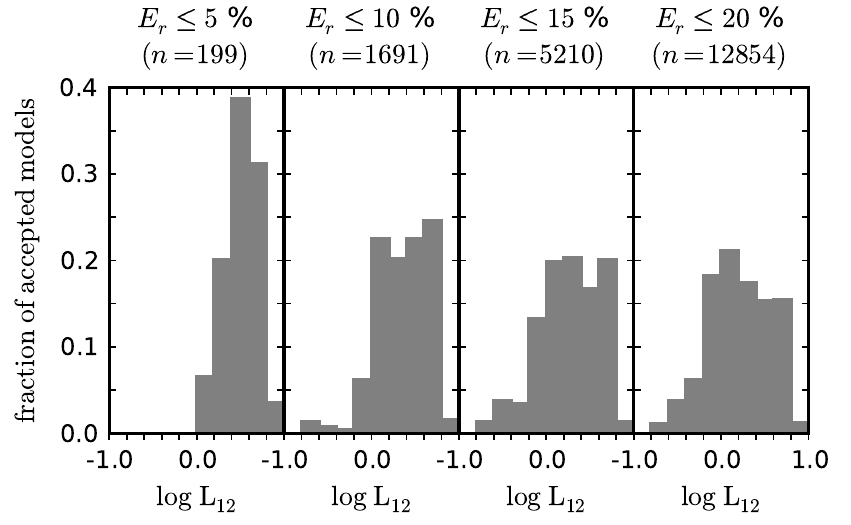}
\caption{Histograms of the logarithm of the AGN bolometric luminosity,
  \mbox{$\Ltw = L_{\rm AGN} / 10^{12} \Lo$}, derived from the scaling
  of each \C\ model (see Equation~\ref{eq:E}). Each panel corresponds
  to a different maximal acceptance error, as marked at the top
  together with the corresponding number of models. Each bin is 0.2
  wide, and its height is the fraction of all accepted models. The
  mean value of \Ltw\ in each panel, from left to right, is 3.45,
  2.71, 2.37 and 2.14.}
\label{fig:f5}
\end{figure}

Since the central engine is obscured in \sst, a direct measurement of
the AGN bolometric luminosity is impossible. However, the bolometric
flux enters directly into the fitting procedure (see
Equation~\ref{eq:E}) as the scale factor that minimizes the error in
matching the model spectral shape with the data. The source luminosity
$L_{AGN}$ is then derived from its luminosity distance $D_L$ =
1.703\,Gpc, obtained from the redshift $z = 0.325$ for standard
cosmological parameters (\mbox{$H_0 = 70\ {\rm km\ s^{-1}\
    Mpc^{-1}}$}, $\Omega_M = 0.3$, flat universe). The best-fit model
has $\Ltw = L_{AGN}/10^{12} \Lo = 3.47$, and Figure~\ref{fig:f5} shows
the distribution of $\log \Ltw$ derived for all fitted models within a
given acceptance error. At the most restrictive level, all the models
fall in the range $1.1 \le \Ltw \le 6.5$ and the mean value is 3.45,
similar to the best-fit model. As the acceptance becomes less
restrictive, the range of accepted models extends to luminosities
lower than \E{12}\,\Lo, but its upper boundary stays unchanged; the
figure panels for $E_r \le 10\%$ and $E_r \le 20\%$ for instance are
very similar except for the presence of more $L < \E{12}\Lo$ models in
the latter. The reason is simple. The luminosity scale factor is $\int
F_\lambda d\lambda$, and as is evident from Figure~\ref{fig:f1}, a
large fraction of the integral is contained at wavelengths that are
missing from the data as they are shorter than the IRAC
measurements. Model SEDs that drop precipitously before the IRAC
points can still produce a small error estimate $E$ by reasonably
fitting all other, longer wavelengths. Such models will be formally
acceptable--but only because the short wavelength region, crucial for
the luminosity determination, is so poorly sampled in the
data. Observations at these short wavelengths will constrain better
the SED, and provide a more accurate determination of \Ltw. With the
current data, our best estimate is $\Ltw \simeq 3$ with a likely range
of 1--7.

Integrating the mid-infrared (MIR) luminosity only, \citet{Lacy+2007b}
find $L_{\rm MIR} = 0.25\cdot\E{12}\,\Lo$. \citet{Richards+2006} show
that the bolometric correction from the mid-infrared is about a factor
of eight. With this correction, the earlier estimate gives \Ltw\
\about\ 4, in good agreement with the detailed \C\ calculations.

%%%%%%%%%%%%%%%%%%%%%%%%%%%%%%%%%%%%%%%%%%%%%%%%%%%%%%%%%%%%%%%
%%%%%%%%%%%%%%%%%%%%%%%%%%%%%%%%%%%%%%%%%%%%%%%%%%%%%%%%%%%%%%%

\section{FEATURE SHAPE AND ORIGIN OF 10-MICRON EMISSION}
\label{sec:peakshift}

After many years in which it remained undetected in type~1 AGN, the
10\mic\ feature was finally discovered in emission in Spitzer
observations \citep{Siebenmorgen+2005,Hao+2005,Sturm+2005}. In
addition, the 18\mic\ feature appears in quite prominent emission. All
three teams noted the large differences with Galactic sources --- the
10\mic\ emission feature in AGN is much broader, and in most cases its
peak seems to be shifted to longer wavelengths, up to \about
11~\mic. Analyzing the feature with the simple approximation
$\kappa_\lambda B_\lambda(T)$ (optically thin emission from dust at
the single temperature $T$), all three teams found significant
differences between the dust absorption coefficient in AGN and the
interstellar medium, suggestive of a different mix of the silicate
components. Significantly, though, the shifts toward longer
wavelengths apparent in {\em emission features} were never reported in
absorption; AGN {\em absorption features} reach their deepest level at
the same wavelengths as Galactic sources, \about 9.8~\mic. The
different behavior of emission and absorption features suggests that
the apparent peculiarities of AGN emission features do not arise from
the dust composition, but rather from radiative transfer effects.

%%%%%%%%%%%%%%%%%%%%%%%%%%%%%%%%%%%%%%%%%%%%%%%%%%%%%%%%%%%%%%%
\begin{figure}
\includegraphics[width=\columnwidth]{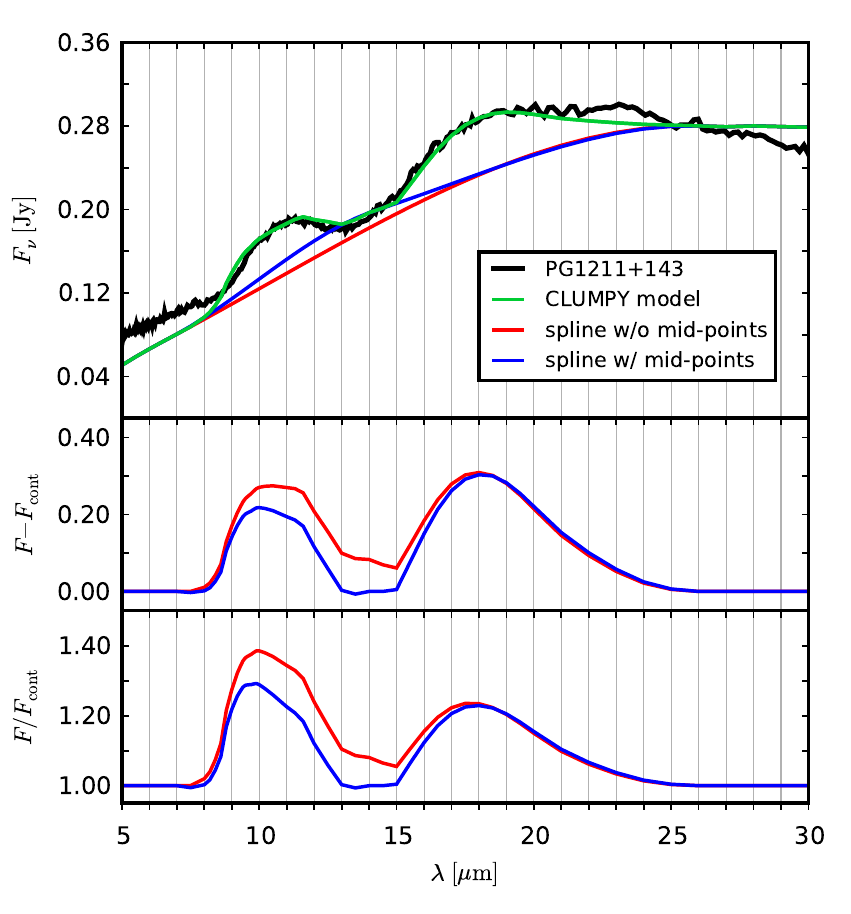}
\caption{Apparent shift of feature peak in quasar PG1211+143 as a
  radiative transfer effect. {\it Top:}~The {\em Spitzer} data (black)
  show the 10\mic\ and 18\mic\ silicate features in emission. The SED
  of the best-fit \C\ model (within 8--30~\mic) is shown in green; its
  parameters are $q = 0$, \mbox{$N_0 = 5$}, $\tv = 20$, $\sigma = 25$,
  $Y = 20$, \mbox{$i = 60$}. The model reproduces observations that
  prompted suggestions for non-standard dust composition. Two
  underlying continua are constructed as splines with (blue) and
  without (red) mid-range pivots over the \mbox{14--14.5~\mic}
  inter-feature region. {\it Middle:}~Continuum-subtracted fluxes for
  each of the continua in the top panel. {\it Bottom:}~The flux ratio
  $F/F_{cont}$ for each continuum.}
\label{fig:f6}
\end{figure}
%%%%%%%%%%%%%%%%%%%%%%%%%%%%%%%%%%%%%%%%%%%%%%%%%%%%%%%%%%%%%%%

%%%%%%%%%%%%%%%%%%%%%%%%%%%%%%%%%%%%%%%%%%%%%%%%%%%%%%%%%%%%%%%
\begin{figure*}
\includegraphics[width=\textwidth]{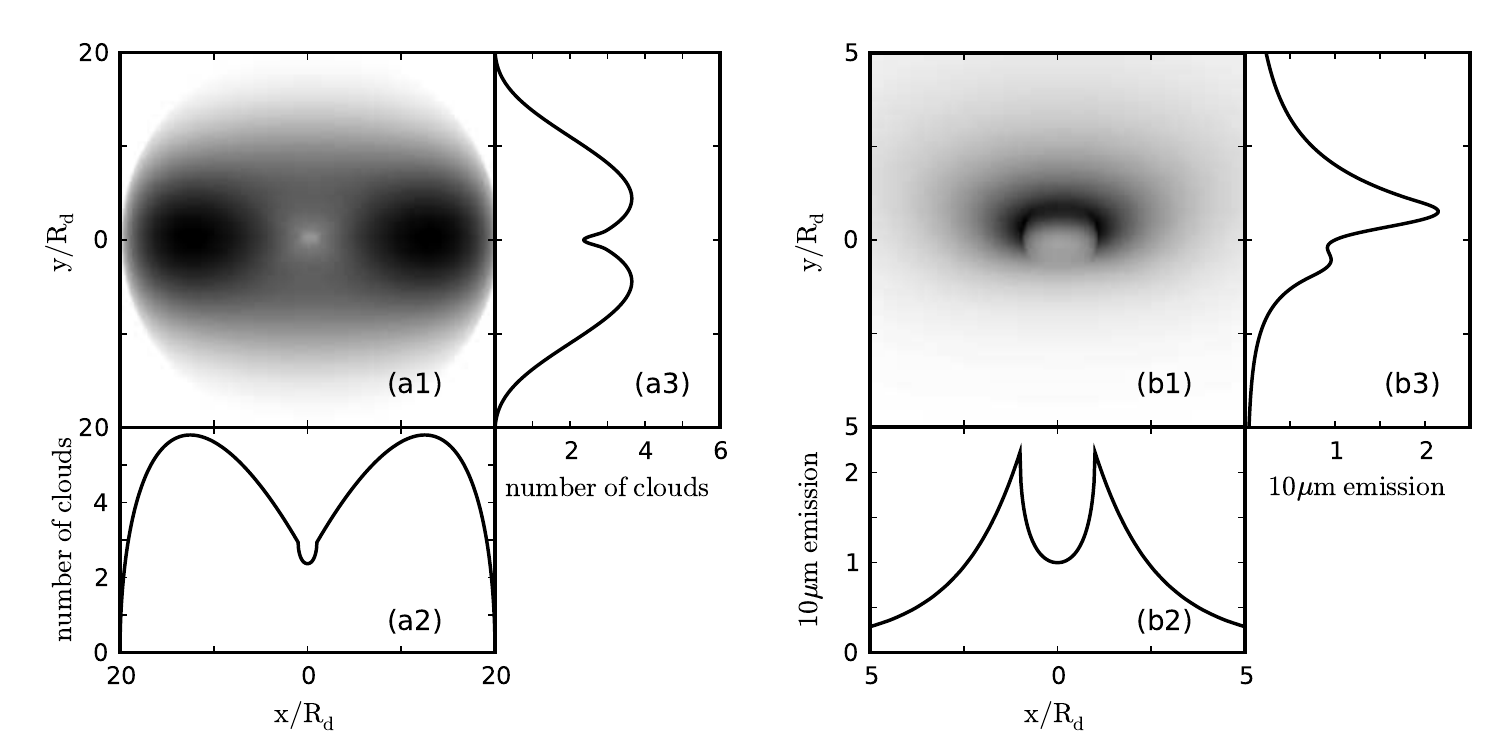}
\caption{Cloud column (left) and 10\mic\ emission (right) along
  viewing rays through the best-fit torus model for \pg\ (see
  Table~\ref{tab:parametervaluesPG1211}). {\it (a1):}~Map of the cloud
  column. Axes are linear displacements $x/R_d$ and $y/R_d$ from the
  central AGN, with $R_d$ the dust sublimation radius. The gray scale
  is linear, with white standing for zero clouds and darker shades
  indicating higher cloud columns. {\it (a2):}~One-dimensional cut
  through the cloud number distribution in (a1) along the
  $x$-coordinate at $y$ = 0. {\it (a3):}~Same as (a2), but vertically
  along the $y$-coordinate at $x$ = 0. {\it (b1):}~The distribution of
  10\mic\ emission emerging from the central $10\,R_d \times
  10\,R_d$. The gray scale is linear, darker shades indicating higher
  10\mic\ emission. {\it (b2):}~The 10\mic\ emission profile along the
  $x$-coordinate at $y=0$, normalized to its central value. {\it
    (b3):}~Same as (b2), but vertically along the $y$-coordinate at
  $x$ = 0.}
\label{fig:f7}
\end{figure*}
%%%%%%%%%%%%%%%%%%%%%%%%%%%%%%%%%%%%%%%%%%%%%%%%%%%%%%%%%%%%%%%

To further investigate this, we analyzed the Spitzer data of quasar
\pg, one of the sources in the original discovery paper of
\cite{Hao+2005}, which shows both silicate features in emission. While
the data cover \mbox{5--35~\mic}, for the model fitting we employed
only wavelengths between 8 and 30~\mic. Fitting the shorter
wavelengths with our torus models proved rather difficult in this
source. The same problem arises in other PG quasars, where
\cite{Mor+2009} find that high flux levels at short wavelengths
necessitate the addition of a hot dust component to the torus emission
in their models \citep[see also][]{Netzer+2007}. Wavelengths longer
than 30~\mic\ were omitted in the fitting because of high noise
levels. The top panel of Figure~\ref{fig:f6} plots the observed SED of
\pg\ between 5 and 30~\mic\ in black color and the SED of the best-fit
model in green. In addition to a prominent 18\mic\ feature, the
displayed model shows a broad 10\mic\ feature that reaches local peak
emission at 11.6~\mic. An analysis of the feature shape requires the
construction of an underlying continuum. \cite{Sirocky+2008} discuss
this problem in detail and show that the proper continuum definition
requires a spline fitted to the two wavelength regions shorter than
the 10\mic\ feature and longer than the 18\mic\ feature, and in
between them. This spline fit to the model results is shown with blue
color in the figure. Although the central region,
\mbox{14--14.5~\mic}, is essential for a correct definition of the
continuum, it was missing from earlier analyses. The corresponding
spline is plotted in red for comparison. The figure middle panel shows
the continuum-subtracted flux in each case. The feature peaks at
10.0~\mic\ in the properly constructed continuum, but has a flat
plateau between \mbox{\about 9.8--11.6~\mic} that peaks nominally at
10.5~\mic\ in the traditional continuum. The bottom panel shows the
ratio $F/\Fcont$ for each continuum. Under the common parametrization
with $\kappa_\lambda B_\lambda(T)$, the $F/\Fcont$ curves would be
taken as the actual dust absorption coefficient. However, they are the
outcome of radiative transfer calculations with the standard
\citetalias{OHM1992} dust, whose absorption profile looks quite
different; these artificial ``absorption coefficients'' are much
flatter than the peaked shape of the input $\kappa_\lambda$.

The reason for the peculiar shape of the emission feature is quite
simple. The feature originates from the optically thin emitting layer
on the bright surfaces of clouds illuminated directly by the
AGN. Absorption by other clouds encountered on the way out toward the
observer alters the feature's shape. This absorption is strongest at
the feature peak, where the absorption coefficient is largest, and
\tv\ \about\ 20 is where single clouds become optically thick at that
peak. When the generated photons encounter \about 1 cloud along the
remaining part of the path toward the observer, the peak is absorbed
while photons in the wings escape freely, effectively flattening the
shape of the feature. An increasing number of clouds along the path
would absorb the peak and the wings of the feature more strongly,
producing a self-absorption dip in the feature's shape, and eventually
suppressing the entire feature (see also Fig.~2 in
\citetalias{Nenkova+2008b}). The apparent shift toward longer
wavelengths arises from the interplay with the shape of the rising
continuum underneath the feature. It may be noted that such apparent
variations in the shape of the silicate emission feature in evolved
stars prompted the suggestion of dust chemical evolution
\citep{Little-MareninLittle1990,Stencel+1990}, but were similarly
shown to reflect radiative transfer effects \citep{IE1995}.

%%%%%%%%%%%%%%%%%%%%%%%%%%%%%%%%%%%%%%%%%%%%%%%%%%%%%%%%%%%%%%%%%%
\begin{deluxetable}{ccccccc}
\tablewidth{\columnwidth}
\tablecaption{Properties of fitted parameters for \pg \label{tab:parametervaluesPG1211}}
\tablehead{\colhead{Parameter} &\colhead{Best Fit} &\colhead{Peak\tk{a}} &\colhead{90\%-Range\tk{b}} &\colhead{B\tk{c}} &\colhead{H\tk{d}} &\colhead{w\tk{e}}}
\startdata
\Y      & 20 & 20               &          20              & 1  & 1.00 & 1.00 \\
\q      & 0  & 0                &   \phn 0 -- 0.5          & 2  & 0.61 & 0.30 \\
\tv     & 20 & 20               &       20 -- 30           & 2  & 0.50 & 0.25 \\
\No     & 5  & 6,7\tk{*}        &   \phn 2 -- 9\phn        & 8  & 0.21 & 0.03 \\
\hline
\sig    & 25 & 25               &       15 -- 60           & 10 & 0.29 & 0.03 \\
\i      & 60 & 0,10,20,40\tk{*} &   \phn 0 -- 70           & 8  & 0.14 & 0.02
\enddata
\tablecomments{Statistical indicators for the sample of all 28 models
  with relative deviation \mbox{$E_r \leq 10\%$} from the best-fit
  model, and with $N_0 \le 10$ (see text), listed in descending order of constraint.
  Footnotes a--f identical to Table~\ref{tab:parametervaluesSST}.}
\tablenotetext{*}{Peak comprises multiple bins; all listed bins have equal
  heights.}
\end{deluxetable}
%%%%%%%%%%%%%%%%%%%%%%%%%%%%%%%%%%%%%%%%%%%%%%%%%%%%%%%%%%%%%%%%%%%%

Table~\ref{tab:parametervaluesPG1211} summarizes the analysis of the
distribution of all \C\ models with fitting errors within 10\% of the
best-fit model. This prescription is identical to the one employed for
\sst, but in the present case it yields only 38 models instead of
1691. These models further break into two distinct groups with
different ranges of \No, the radial number of clouds in the equatorial
plane. While 28 models have $N_0 \le 9$, the other 10 fall in the \No\
= 16--18 range, with a large gap between the two groups. Because
values of \No\ larger than \about 10 are unlikely in general (see
\S3.4 in \citetalias{Nenkova+2008b}), we exclude the ten models with
$N_0 \ge 16$ from our sample.

As before, only three parameters are well constrained. The radial
cloud distribution is again well-constrained, but now it is flat with
$q = 0$. This leads to a strongly constrained torus thickness $Y =
20$, in sharp contrast with \sst\ where \Y\ is the least-well
constrained parameter. All 28 models in the selected sample have the
same value of \Y, although this probably reflects our discrete
sampling of parameter space; there could be a small range around \Y\ =
20, but 10 and 30 are clearly excluded. The cloud optical depth is
well-constrained at $\tv \approx 20-30$. On the other hand, while well
constrained for \sst, \No\ is the least well constrained parameter
here, with a likely range of \mbox{2--9} clouds.

As noted above, flat-top emission features arise from absorption by a
single cloud with \tv\ \about\ 20. As is evident from
Table~\ref{tab:parametervaluesPG1211}, all accepted models have \tv\
\about\ 20--30, but \No\ is largely unconstrained. However, the
average number of clouds along the line of sight to the AGN (see
Equation~\ref{eq:Ni}) falls within the narrow range $0 < N(i) \le 2.5$
for all accepted models. To a certain degree, $N(i)$ is a good proxy
for the typical number of clouds along lines of sight that pass close
to the dust sublimation radius, where the 10\mic\ emission is
originating. In panel (a1) of Figure~\ref{fig:f7} we show the number
of clouds along all lines of sight through the best-fit torus model.
Panel (a2) shows a one-dimensional cut through the image, providing
the profile of the number of clouds per ray along the $x$-coordinate
at $y=0$. Due to the axial symmetry of the torus, this profile is
symmetric with respect to $x=0$, irrespective of the viewing
angle. The signature of the central cavity is clearly visible in this
profile: the cloud column reaches a minimum of 2.4 at the center and
stays close to this level for all $x/R_d \le 1$. It reaches a maximum
of $\approx 6$ clouds along rays roughly $10\,R_d$ away from the AGN.
Panel (a3) shows the corresponding profile in the vertical direction
at $x = 0$. The symmetry of this profile around $y=0$ again reflects
the axial symmetry, which ensures equal path lengths through the torus
above and below the central line of sight.

While the number of clouds along two lines of sight displaced
symmetrically from the center is equal, the illumination patterns of
individual clouds as seen by the observer can differ for the two,
depending on the position angle in the plane of the sky. In panel (b1)
we plot the two-dimensional distribution of the model 10\mic\ emission
for the central region with size $10\,R_d \times 10\,R_d$. Roughly
50\% of the flux is detected within the inner $5\,R_d$ radius, and
70\% of that fraction comes from the image upper half. This radiation
originates from regions on the far inner face of the torus; no
emission originates from the near side, where the observer faces the
dark sides of the clouds. Similar to panels (a2) and (a3), we plot in
panels (b2) and (b3) profiles of the 10\mic\ emission along the $x$
and $y$ directions. As expected, the horizontal profile in panel (b2)
is symmetrical, clearly displaying the dust-free cavity at its
center. On the other hand, the shape of the vertical profile in (b3)
reveals the asymmetry between the emission in the upper and lower
halves. Despite equal cloud columns along viewing lines above and
below the image center, the emission is not equal, owing to the strong
anisotropy of single cloud emission. The 10\mic\ emission originates
from hot, bright surfaces of clouds located on the torus far inner
face. Clouds in the torus near side, which show their dark, cooler
faces, only absorb the 10\mic\ photons that were emitted on the torus
far side.

Most clumpy torus models do not produce the apparent shift in peak
emission. The shifts occur predominantly in models that have a small
\tv\ ($\la$ 20). Significantly, \tv\ \about\ 20 models are also the
ones producing the most prominent 10\mic\ emission features across the
likely range of \tv. As is evident from Fig.~16 in
\citetalias{Nenkova+2008b}, the emission feature strength decreases
monotonically as \tv\ increases up to \tv\ \about 70; in some cases
the feature even switches to absorption for pole-on viewing. Therefore
low-\tv\ models stand out in their feature strength and it is
reasonable that such sources would be preferentially selected in
observations that looked to identify the 10\mic\ silicate emission
feature in AGN. Finally, it should be noted that the absorption
coefficients widely used in the literature do not have their peaks at
9.8~\mic. In the tabulation of \cite{Draine2003}, the feature peaks at
9.48~\mic\ instead. The ``cold'' silicate dust of
\citetalias{OHM1992}, which is the one used here, has its peak at
10.0~\mic. The effect on the present discussion is insignificant.

\section{SPECTRAL PROPERTIES OF CLUMPY MODELS}
\label{sec:s10}

In addition to the detailed fitting of the \sst\ and \pg\ data, we
investigated some other properties of the 10\mic\ feature, comparing
observations with general properties displayed by the model database.

\cite{Hao+2007} present a large compilation of {\em Spitzer} mid-IR
spectra. Although a loosely defined sample, it is the largest gathered
thus far. For each source they measure the 10\mic\ silicate feature
strength $S_{10}$ from
\begin{equation}
  \label{eq:s10}
  S_{10} = \ln \frac{F(\lambda_{10})}{\Fcont(\lambda_{10})}\,,
\end{equation}
where $F$ is the measured flux, \Fcont\ is a continuum constructed
underneath the 10\mic\ and 18\mic\ silicate features \citep[see][for
details; see also \S\ref{sec:peakshift}]{Sirocky+2008}, and
$\lambda_{10}$ is the peak wavelength of the feature strength;
emission features have a positive $S_{10}$, absorption features a
negative one.\footnote{The feature strengths of \sst\ are $S_{10}$ =
  0.26 and $S_{18}$ = 0.34 (for the 18\mic\ feature). Both are
  uncertain to within \about\ $\pm$ 0.1. For the model shown in
  Figure~\ref{fig:f6}, the feature strengths determined from the blue
  curve are $S_{10} = 0.26$ and $S_{18} = 0.21$ (at 18.0~\mic), and
  $S_{10} = 0.33$ and $S_{18} = 0.21$ (at 17.5~\mic) from the red
  curve.} It may be noted that the specific prescription of continuum
construction modifies, and can even reverse, the relative strengths of
the 10\mic\ and 18\mic\ features, as is also apparent from
Figure~\ref{fig:f6}; the ratio of the two strengths is an important
indicator of dust optical properties \citep{Sirocky+2008}.

%%%%%%%%%%%%%%%%%%%%%%%%%%%%%%%%%%%%%%%%%%%%%%%%%%%%%%%%%%%%%%%
\begin{figure}
\includegraphics[width=\columnwidth]{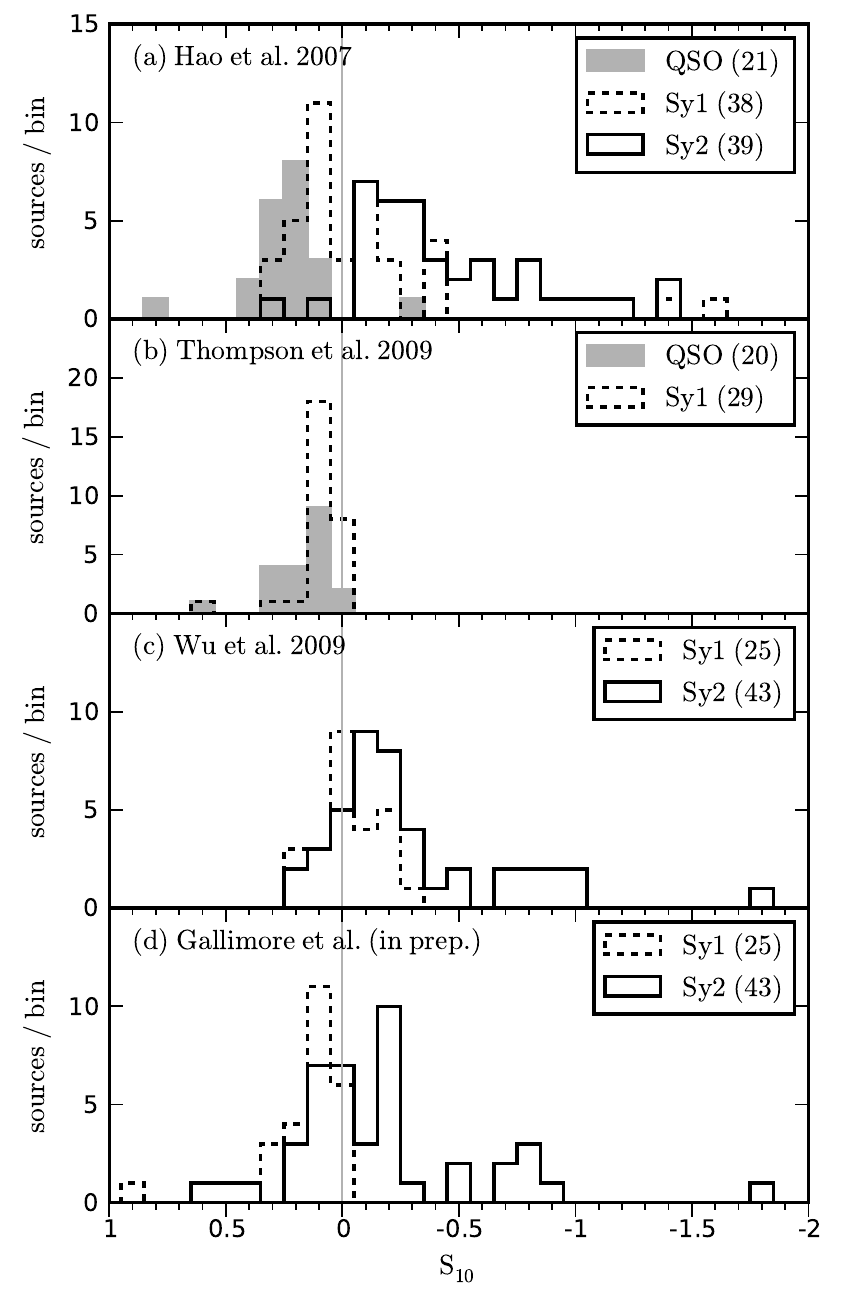}
\caption{Distributions of the 10\mic\ silicate feature strength \sten\
  (Equation~\ref{eq:s10}) in several AGN samples. The bin size is
  0.1. All ULIRGs present in the original samples have been
  removed. Measurements for QSOs are shown with gray bars, Seyfert~1s
  with dashed lines, and Seyfert~2s with solid lines. The number of
  sources of different type is given in parentheses in the
  legend. {\it (a):}~\emph{Spitzer} sample by \citet{Hao+2007}. {\it
    (b):}~Archival sample of type~1 sources by
  \citet{Thompson+2009}. Note the different scale. {\it (c):}~Seyfert
  sources from the 12\mic\ Galaxy Sample, presented by
  \citet{Wu+2009}, and {\it (d)} re-analyzed by Gallimore et~al. (in
  preparation). Panel (c) contains only sources also present in (d),
  and the source classification in both panels is adopted from the
  latter.}
\label{fig:f8}
\end{figure}
%%%%%%%%%%%%%%%%%%%%%%%%%%%%%%%%%%%%%%%%%%%%%%%%%%%%%%%%%%%%%%%

Removing all ULIRGs, the \citet{Hao+2007} sample contains 21 QSOs, 38
Seyfert~1 and 39 Seyfert~2 galaxies. The top panel of
Figure~\ref{fig:f8} shows the histograms of the feature strengths for
the three groups. The figure other panels show results from recent
studies, which produced additional compilations of feature strengths:
\cite{Thompson+2009} compared a sample of Seyfert~1 galaxies with
quasars, \cite{Wu+2009} and Gallimore et~al. (in preparation) analyzed
Seyfert galaxies, both type~1 and~2, from the 12\mic\ Galaxy Sample
\citep{Rush+1993}. While Wu et~al.\ adopt the original source
classification of Rush et~al., Gallimore et~al.\ establish a different
source type in several cases, based on work published elsewhere. We
employ the latter classification in both panels (c) and (d),
dispensing with all sources re-classified as LINERs or HII
(star-forming) galaxies and confirming the Wu et~al.\ suggestion that
the re-classification of several sources has little effect on the
statistical results of \sten\ measurements.

In addition to the torus emission, the infrared radiation of many
active galaxies contains a starburst contribution whose fractional
strength varies from source to source
\citep[e.g.,][]{Netzer+2007}. Removing the starburst component by
subtracting a suitable template and leaving no PAH residuals is thus
an important preliminary step in the detailed SED analysis of many
individual AGN \citep*[see, e.g.,][]{Mor+2009}. Note, however, that
the two sources analyzed here in detail show no signs of ongoing star
formation, either in the form of PAH emission or far-IR (FIR)
emission. While PAH emission can contaminate the 10\mic\ region in
some individual spectra, its overall impact on the averages of large
samples seems minimal. \citet{Netzer+2007} subtract a starburst
template from the average spectra of AGN with and without strong FIR
detections and find that the MIR regions are hardly affected by this
subtraction in either case. In particular, their Figure~6 shows that
the strength of the 10\mic\ silicate feature barely changes. The
analyses by \cite{Wu+2009} and Gallimore et~al.\ (in preparation) of
the same data set provide an even stronger evidence: The former
ignores the potential starburst contribution while the latter includes
a PAH component, handled with the \textsc{Pahfit} tool
\citep{Smith+2007}. In spite of this difference, the histograms in
panels (c) and (d) of Figure~\ref{fig:f8} are quite similar, showing
comparable lower and slightly increased upper limits on \sten\ and an
overall shape that is essentially the same.

Comparison of the histograms for type~1 sources in the panels of
Figure~\ref{fig:f8} shows that in moving from Seyfert to quasar
luminosities the 10\mic\ feature shifts to enhanced emission. This
trend was noted earlier in \citeauthor{Nenkova+2008b}
(\citeyear{Nenkova+2008b}; see \S6.4), and the analysis here verifies
this suggestion, giving it quantitative evidence. Nenkova et~al.\
point out that the most likely explanation is that the number of
clouds along radial rays is smaller in quasars than in Seyferts.

\begin{figure}
\includegraphics[width=\columnwidth]{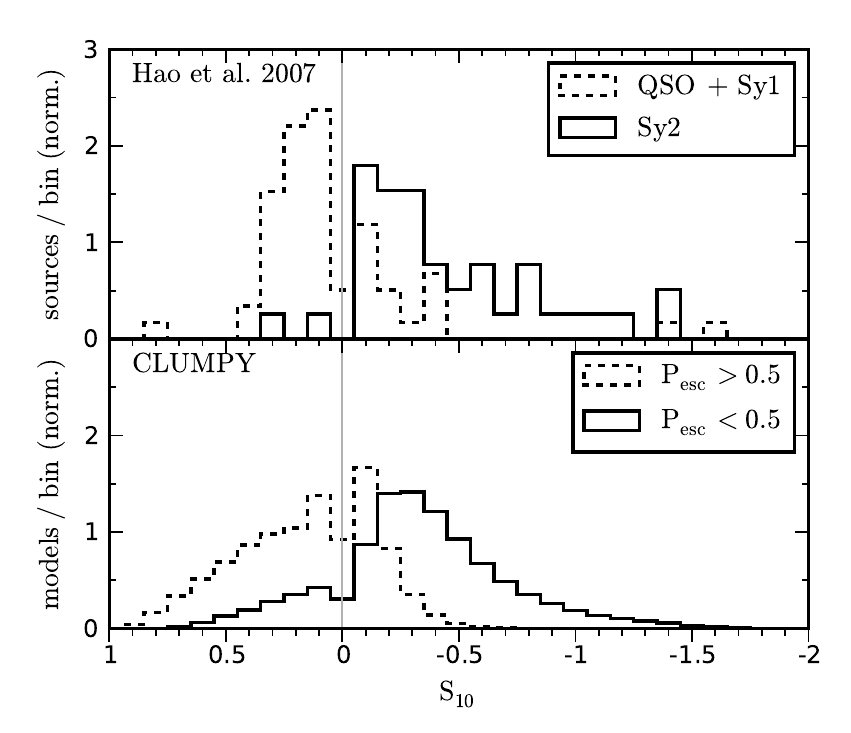}
\caption{Distributions of the 10\mic\ silicate feature strength
  \sten. The bin size is 0.1 and each histogram is normalized to unit
  area. {\it Top:}~Data from the \citet{Hao+2007} sample. Dotted line
  shows type~1 sources (QSO and Seyfert~1 combined), solid line
  Seyfert~2s. {\it Bottom:}~Histograms for 840,000 \C\ models whose
  parameters most likely correspond to physical values (see
  text). Models with escape probability $\Pesc > 0.5$ (likely type~1
  source in a clumpy torus) are shown as dotted line, those with
  $\Pesc < 0.5$ (likely type~2) as solid line. For statistical
  properties of all samples see Table~\ref{tab:s10}.}
\label{fig:f9}
\end{figure}

Grouping together the QSOs and Seyfert~1s of the \citet{Hao+2007}
sample, the top panel of Figure~\ref{fig:f9} shows the histograms and
Table~\ref{tab:s10} lists the statistical indicators of the \sten\
distributions in type~1 and~2 sources. Most sources exhibit rather
small absolute values of \sten. The histogram of type~1 sources is
clearly shifted toward emission in comparison with type~2. Although
the \citet{Hao+2007} sources do not constitute a complete sample, the
selection criteria were unrelated to the silicate feature. The derived
histograms can thus be reasonably considered representative of the
differences between types~1 and~2.

Our clumpy torus models should produce similar histograms if they bear
a resemblance to the IR emission from AGN. Such a comparison presents
two fundamental difficulties. First, the assignment of a given clumpy
model to type~1 or~2 is not deterministic --- only a probability can
be assigned. We handle this problem by dividing the models according
to the probability \Pesc\ for an unobscured view of the AGN. The
collection of models with $\Pesc > 0.5$ can be expected to resemble
the behavior of the type~1 population, those with \mbox{\Pesc\ $<
  0.5$} type~2. The second problem is that the actual distribution of
parameter values is unknown. Since we do not have any handle on these
distributions, we decided to test the adequacy of histograms produced
by a uniform sampling of the model parameters within the bounds
deduced in \citetalias{Nenkova+2008b}: $0 \le q \le 3$, $N_0 \le 15$,
$30 \le \tv \le 100$, $15\deg \le \sigma \le 60\deg$, and $10 \le Y
\le 100$. Since \q\ was sampled here more thoroughly than in
\citetalias{Nenkova+2008b}, we use the full range listed in
Table~\ref{tab:clumpyparams}. The parameters were sampled in steps of
0.5, 1, 10, 5, 10, 10 for \q, \No, \tv, \sig, \Y, \i,
respectively. The bottom panel of Figure~\ref{fig:f9} shows the
histograms of \sten\ for all database models selected by these
criteria. These distributions resemble those of the observational
sample, as is also evident from their statistical properties listed in
Table~\ref{tab:s10}. The $1\sigma_s$ and $2\sigma_s$ ranges of \sten\
given in the table contain sources and models within 1 and 2 standard
deviations $\sigma_s$ from the mean of each of the two
distributions. Leading to exclusion of only few sources and models at
the very ends of the distributions, this additional selection has the
effect of a much more meaningful agreement of the two distribution
widths, not spoiled by rare outliers. At the $2 \sigma_s$ level,
rejected sources are just 3 type~1 and 2 Seyfert~2s, and among the
sample of models only 4\% of those with \mbox{\Pesc\ >\ 0.5} and 6\%
with \mbox{\Pesc\ <\ 0.5} are excluded due to this criterion.

Although the choice of uniform sampling of the model database is
arbitrary, it produces reasonable results. The reason is that, as
noted already in \citetalias{Nenkova+2008a} and
\citetalias{Nenkova+2008b}, clumpy models never produce very deep
absorption features, in agreement with observations. This limited
range is reflected in the histograms for any reasonable criteria used
for model selection from the database. The other main characteristic
of the observed histograms is the separation between type~1 and~2
sources, and this, too, is reproduced reasonably well by the models.

%%%%%%%%%%%%%%%%%%%%%%%%%%%%%%%%%%%%%%%%%%%%%%%%%%%%%%%%%%%%%%%
%%%%%%%%%%%%%%%%%%%%%%%%%%%%%%%%%%%%%%%%%%%%%%%%%%%%%%%%%%%%%%%

%%%%%%%%%%%%%%%%%%%%%%%%%%%%%%%%%%%%%%%%%%%%%%%%%%%%%%%%%%%%%%%
% with original data from Lei Hao
\begin{deluxetable}{ccccc}
\tablecolumns{5}
\tablewidth{\columnwidth}
\tablecaption{$S_{10}$ statistics\label{tab:s10}}
\tablehead{\colhead{}   & \multicolumn{2}{c}{Hao et al. (2007)} & \multicolumn{2}{c}{\textsc{Clumpy}}\\
source type             & QSO\ +\ Sy1 & Sy2                     & $\Pesc > 0.5$ & $\Pesc < 0.5$ \\
sample size             & 59          & 39                      & 340,000       & 500,000}
\startdata
mean                    & 0.03        & -0.46                   &  0.15         & -0.33 \\
median                  & 0.12        & -0.34                   &  0.12         & -0.32 \\
$\sigma_s$              & 0.36        & \phm{-}0.40             &  0.29         & \phm{-}0.38 \\
1$\sigma_s$-range\tk{a} & -0.30, 0.37 & -0.83, -0.06            & -0.14, 0.44   & -0.71, 0.04 \\
2$\sigma_s$-range\tk{b} & -0.44, 0.40 & -1.20, \phm{-}0.29      & -0.42, 0.73   & -1.09, 0.42
\enddata
\tablecomments{Samples of AGN and of \C\ models as in Figure~\ref{fig:f9}.}
\tablenotetext{a}{Ranges of 1 standard deviation $\sigma_s$ from a sample's mean value.}
\tablenotetext{b}{The 2-standard deviations range.}
\end{deluxetable}
%%%%%%%%%%%%%%%%%%%%%%%%%%%%%%%%%%%%%%%%%%%%%%%%%%%%%%%%%%%%%%%

\section{SUMMARY AND DISCUSSION}
\label{sec:summary}

Spitzer IR observations of AGNs have increased significantly the
number and quality of SEDs for these objects and produced some
puzzling results, especially with regard to the 10\mic\ silicate
feature. These include (1) detection of the feature in emission in
type~2 sources, (2) emission features with broad, flat-topped peaks
shifted toward long wavelengths in several type~1 sources, and (3)
absence of any deeply absorbed features. None of these observations
can be satisfactorily explained with smooth density torus models.

Here we have shown that clumpy torus models provide reasonable
explanations for all three puzzles. To that end we have fitted the
Spitzer SEDs of two very different sources with our \C\ models. One
source, \sst, is the first type~2 QSO to show a clear 10\mic\ emission
feature. Our analysis provides a reasonable fit of the SED with a
model that shows the feature in emission. In contrast with smooth
density models, where the AGN is either obscured or visible, our model
produces a small obscuration probability, \Pobs\ = 27\%, for this type
2 source. This relatively low probability may explain why \sst\ is the
only source among more than 20 type 2 QSOs with measured SEDs
\citep[see, e.g.,][]{Polletta+2008} to show a clear 10\mic\ emission
feature.

Addressing the second puzzle, \pg\ is one of the first QSOs to display
the 10\mic\ silicate feature in emission, a feature that is
unexpectedly broad and apparently shifted to longer wavelengths. The
original attempts to explain these properties invoked non-standard
chemical dust composition. Our modeling shows that the shifts are only
apparent and result from the flattening of the feature peak by
radiative transfer in clumpy media. The feature is well reproduced by
clumpy models with standard dust. The third observational puzzle, lack
of deep 10\mic\ absorption features in any AGN, has already been shown
to be a signature of clumpy dust distributions \citep{Nenkova+2002,
  Levenson+2007, Sirocky+2008, Nenkova+2008a, Nenkova+2008b}. Here we
go a step further and produce the histogram of 10\mic\ feature
strength for a large sample of AGN \C\ models. The result is in good
qualitative agreement with the sample observed by \cite{Hao+2007}. In
particular, the median values of both type~1 and type~2 observed
distributions and their widths are well reproduced by the model
database.

The IR SED generally does not constrain very tightly the properties of
dusty sources --- the large degeneracy of the radiative transfer
problem for heated dust is well known
\citep[e.g.,][]{Vinkovic+2003}. In the present case, the problem is
further exacerbated by the clumpy nature of the dust distribution and
the non-spherical geometry. Our model database contains close to 5
million entries, and although the fitting procedure eliminates most of
them, many produce reasonable agreement with the observations. In the
case of \sst, close to 1,700 models with very different parameters
deviate by no more than 10\% from the best-fit model. And while the
much higher quality of data in \pg\ greatly reduces the number of
acceptable models, there are still 28 different ones that are
practically indistinguishable in the quality of their fits. In the
face of this degeneracy, we have developed a statistical approach to
assess the meaningfulness of the various torus parameters derived from
the fits. We find that some parameters are well constrained in each
case, while others are not. In both sources the power law of the
radial distribution (\q) and the optical depth of a single cloud (\tv)
are well constrained, while the torus viewing angle (\i) and its
angular thickness (\sig) are not. Both the cloud number (\No) and
radial thickness (\Y) are well constrained in only one of the sources,
a different one in each case. \citet{AsensioRamos2009} have recently
developed a different, novel approach to tackle the degeneracy
problem. They interpolate the \C\ SEDs by means of an artificial
neural network function, allowing them to study the parameter
distributions as if they were continuous, and employ Bayesian
inference to determine the most likely set of parameters. Applying
this method to a selection of sources, \citet{RamosAlmeida+2009} find
that the principal ability to constrain different \C\ parameters
strongly depends on the individual source. We have already begun an
extensive comparison of the two approaches and will report our
findings elsewhere.

Although the SED alone is generally insufficient for determining all
the torus parameters with certainty, the success in resolving
outstanding puzzling behavior of the 10\mic\ feature in AGN is
encouraging and enhances confidence in the clumpy torus paradigm.

\section*{ACKNOWLEDGMENTS}

We are indebted to Jack Gallimore for kindly providing his
measurements of silicate feature strengths prior to publication and
agreeing to their inclusion in Figure~\ref{fig:f8}. We thank Lei Hao
for providing her original data (included in Figures~\ref{fig:f8} and
\ref{fig:f9} and Table~\ref{tab:s10}), and Nancy Levenson for the SED
of \pg\ (Figure~\ref{fig:f6}). We further thank Cristina Ramos
Almeida, Eckhard Sturm and the anonymous referee for valuable
comments. ME acknowledges the support of NSF (AST-0807417) and NASA
(SSC-40095). RN appreciates early PhD support by A. Feldmeier and DFG
(Fe 573/3).

\bibliographystyle{apj}
%\bibliography{/home/robert/papers/references}

\end{document}